\documentclass[aos]{imsart}

\RequirePackage[OT1]{fontenc}
\RequirePackage{amsthm,amsmath,graphicx,multirow}
\RequirePackage[numbers]{natbib}
\RequirePackage[colorlinks,citecolor=blue,urlcolor=blue]{hyperref}
\usepackage[dvipsnames]{xcolor}

\arxiv{arXiv:0000.0000}

\usepackage{booktabs}
\usepackage{multirow}
\usepackage{color}

\usepackage{latexsym}
\usepackage{amsmath}
\usepackage{amssymb}
\usepackage{amsfonts}
\usepackage{bm}

\DeclareMathOperator*{\argmax}{arg\,max}
\DeclareMathOperator*{\argmin}{arg\,min}

\newcommand{\lp}{\left(}
\newcommand{\rp}{\right)}
\newcommand{\llp}{\left\{}
\newcommand{\rrp}{\right\}}
\newcommand{\lllp}{\left[}
\newcommand{\rrrp}{\right]}
\newcommand{\pd}{\partial}
\newcommand{\eff}{\mathrm{eff}}

\newcommand{\Hk}{\mathcal{H}_k}

\newcommand{\HT}{\mathrm{HT}}

\def\ba#1\ea{\begin{align*}#1\end{align*}} 
\def\banum#1\eanum{\begin{align}#1\end{align}} 

\usepackage{scalerel,stackengine}
\stackMath
\newcommand\reallywidehat[1]{%
\savestack{\tmpbox}{\stretchto{%
  \scaleto{%
    \scalerel*[\widthof{\ensuremath{#1}}]{\kern-.6pt\bigwedge\kern-.6pt}%
    {\rule[-\textheight/2]{1ex}{\textheight}}
  }{\textheight}%
}{0.5ex}}%
\stackon[1pt]{#1}{\tmpbox}%
}

\startlocaldefs

\numberwithin{equation}{section}

\theoremstyle{plain}
\newtheorem{cor}{Corollary}[section]

\newtheorem{thm}{Theorem}[section]
\newtheorem{lem}{Lemma}[section]
\newtheorem{rem}{Remark}[section]

\newtheorem{example}{Example}

\endlocaldefs

\begin{document}

\begin{frontmatter}
\title{Semiparametric adaptive estimation under informative sampling}
\runtitle{Semiparametric adaptive estimation under informative sampling}

\begin{aug}
\author[A]{\fnms{Kosuke}~\snm{Morikawa}\ead[label=e1]{k.morikawa.es@osaka-u.ac.jp}\orcid{0000-0002-6021-5180}},
\author[A, B]{\fnms{Yoshikazu}~\snm{Terada}\ead[label=e3]{yoshikazu.terada.es@osaka-u.ac.jp}\orcid{0000-0002-4509-1108}}
\and
\author[C]{\fnms{Jae Kwang}~\snm{Kim}\ead[label=e2]{jkim@iastate.edu}\orcid{0000-0002-0246-6029}}
\address[A]{Graduate School of Engineering Science,
Osaka University\printead[presep={,\ }]{e1,e3}}

\address[B]{Center for Advanced Integrated Intelligence Research,
RIKEN\printead[presep={,\ }]{}}

\address[C]{Department of Statistics,
Iowa State University\printead[presep={,\ }]{e2}}
\end{aug}

\begin{abstract}
In probability sampling, sampling weights are often used to remove selection bias in the sample. The Horvitz-Thompson estimator is well-known to be consistent and asymptotically normally distributed; however, it is not necessarily efficient. This study derives the semiparametric efficiency bound for various target parameters by considering the survey weights as random variables and consequently proposes two semiparametric estimators with working models on the survey weights. One estimator assumes a reasonable parametric working model, but the other estimator does not require specific working models by using the debiased/double machine learning method. The proposed estimators are consistent, asymptotically normal, and can be efficient in a class of regular and asymptotically linear estimators. A limited simulation study is conducted to investigate the finite sample performance of the proposed method. The proposed method is applied to the 1999 Canadian Workplace and Employee Survey data.

\end{abstract}

\begin{keyword}[class=MSC]
\kwd[Primary ]{62F35}
\kwd{62G20}
\kwd[; secondary ]{62G10}
\end{keyword}
\begin{keyword}
\kwd{double/debiased machine learning} 
\kwd{semiparametric efficency bound}
\kwd{weight model}
\end{keyword}

\end{frontmatter}

\section{Introduction}
\label{sec:1}

Probability sampling is a classic tool for obtaining a representative sample from a target population.  Although probability sampling is a gold standard for estimating descriptive parameters, such as population totals or proportions, it can also be used for analytic inference, which offers answers to scientific queries.  In analytic inference, we are mainly interested in estimating the parameters in the superpopulation model that generates the target finite population.   \citet{korn1999}, \citet{Chambers2003} and \citet{heeringa2010} provide overviews of the analytic inference in survey sampling.

Survey sample data enable  inferences about  superpopulation models without the need to  observe every element in the finite population. In probability sampling, first-order inclusion probabilities are known.  
If the sampling weights are correlated with the study outcome variables even after adjusting for the covariates, the sampling design is called informative. Informative sampling occurs when there is a discrepancy between design variables and auxiliary  variables used for regression analysis, notably even in widely utilized methods such as Poisson sampling and probability proportional to size sampling \citep{Slud2020}.  
\citet{Pfeffermann1999} 
presented likelihood-based methods for analytic inference under informative sampling. While  survey weights contribute to the development of a design-based estimator with statistically desirable properties, such as consistency and asymptotic normality, they do not guarantee efficiency. Addressing  how to improve efficiency while ensuring consistency under informative sampling is a fundamental problem in survey sampling.  

In the context of regression analysis, 
\citet{magee98} showed that consistency is retained under any multiplication of the weights by a function of covariates and suggested ways to facilitate the selection of such a function. \citet{Pfeffermann1999} proposed a modification of the weights as a function of the covariates. \citet[Section 6.3.2]{fuller09} provided suggestions to maximize the efficiency of a class of estimators using modified weights. 


More generally, modifying the design weight directly to improve efficiency while maintaining consistency is another important direction of research in analytic inference.  
\citet{beaumont08} showed that replacing the weights with smoothed weights, which is the conditional expectation of the weights given the sampled data, could produce a more efficient estimator. 
In the estimation of linear regression models, \citet[Chap.6]{fuller09} proposed the optimal weight function under non-informative sampling, while \citet{kim13} proposed an optimal weight function for generalized linear models under informative sampling. These weights are optimal in the sense that the asymptotic variance of the estimator with weighting methods is the smallest. 

In this article, our goal is to broaden the class of estimators that can be considered optimal under the semiparametric theory framework, extending beyond the conventional focus on linear regression models. 
Under these semiparametric models, an optimal estimation in the class of regular and asymptotically linear (RAL) estimators has been the research hotspot in mainstream statistics \citep{bickel1998, van2000asymptotic}. Consequently, the optimal estimation technique under the semiparametric model has been developed in many application areas, for example, in missing data \citep{robins1995analysis, rotnitzky1998semiparametric, morikawa2021, zhao2021}, measurement error models \citep{ma2006}, and longitudinal data analysis \citep{li2011}.
However, to our best knowledge,  efficient estimation in informative sampling other than linear regression models has not been explored in the literature.
Existing work on efficient estimation under survey sampling,  such as Isaki and Fuller \citep{isaki1982} and Han and Wellner \citep{han2021},  have been developed only under noninformative sampling. The assumption of noninformative sampling  is indeed a strong assumption and may not hold in many real situations.   Therefore, this is an important research gap that needs to be addressed. 

Our contribution can be summarized as follows. First, in the context of informative sampling, we have developed a unified approach to semiparametric efficient estimation using  the projection  theory \citep{bickel1998, tsiatis2006}. This approach is versatile,   applicable to a wide array of target parameters including  the population mean, the parameter in the regression model, and the parameter in the parametric outcome model. 
Our proposed method, therefore, has broad applicability across numerous survey sampling scenarios where sampling weights are known.  
Second, we have introduced adaptive estimators that asymptotically attain the semiparametric efficiency bound, rendering them asymptotically optimal within the extensive class of Regular Asymptotically Linear (RAL) estimators for the parameters under consideration. By innovatively incorporating models on the sampling weights, our estimators utilize these weights more effectively, enhancing the estimation accuracy of various parameters. This approach marks a departure from traditional descriptive inference by treating weights as random variables in analytic inference, thereby leveraging the weight model to improve the efficiency of parameter estimation.
Third, to mitigate bias arising from potential  misspecification of weight models, we have developed a nonparametric estimation of the weight models. Utilizing the debiased/double machine learning method \citep{chernozhukov2018double},  this innovation is applied to our semiparametric adaptive estimators, enhancing their robustness. Consequently, our methodology stands out not only for its efficiency,  but also for its robustness, thanks to the nonparametric estimation of the weight model. 
Moreover, we have rigorously established the large-sample properties of the adaptive efficient estimator, including
$\sqrt{n}$-consistency and asymptotic normality. As a result, this foundation enables statistical inference using confidence intervals, providing a scientific framework for assessing the precision and reliability of estimates derived from large samples.

An extensive simulation study is performed to evaluate the performance of the proposed methods. 
We also present an application of the proposed estimator to the Canadian Workplace and Employee Survey data \citep{patak98, fuller09}.


\section{Basic setup}
\label{sec:2}
\subsection{Notation}
\label{sec:2.1}
Suppose that a finite population $(X_i, Y_i, W_i)~(i=1, \ldots, N)$ identically and independently follows an unknown distribution, where $X_i$ is a covariate, $Y_i$ is a response variable, and $W_i$ is the sampling weight associated with unit $i$. We denote our target parameter by $\theta$, which is a parameter in the distribution of $X$ and $Y$, such as the coefficients in a parametric regression model $E(Y\mid x;\theta)$. A finite sample of size $n$ is obtained by the Poisson sampling, i.e., the sampling indicators $\delta_i\,(i=1,\dots, N)$ are independently distributed as $\delta_i\mid (W_i=w_i)\sim Bernoulli(w^{-1}_i)\;(i=1, \ldots, N)$, where $\delta_i$ takes the value 1 if the $i$-th unit is sampled and 0 otherwise, and $W^{-1}_i$ is called the first inclusion probability. The sampling weight $W_i$ can be correlated with the study variable $Y_i$ conditional on $X_i$, which is often called informative sampling. On the other hand, \citet{han2021} assumed that $Y_i$ is independent of $\delta_i$ given $X_i$, which corresponds to noninformative sampling. This study considers two settings: $X$ is available for (i) all units in the finite population; (ii) units only with $\delta_i=1$. 
It should be noted that even when additional variables 
$Z$, which may correlate with $W$ and $Y$, are accessible, incorporating $Z$ does not improve the efficiency of estimating $\theta$. In scenarios involving non-probability sampling or missing data analysis, where $W$ is unobservable, 
$Z$ becomes essential to account for $\delta$. However, in the context of probability sampling, this effect is blocked by the presence of $W$. Consequently, the information offered by
$Z$ becomes redundant and can be safely ignored.




Regarding the parameter $\theta$ of interest, we  consider three situations that are useful in many applications: (a) the unique solution to $E\{U(\theta; X, Y)\}=0$; (b) a parameter of the regression model $E(Y\mid x; \theta)=\mu(x;\theta)$; (c) a parameter of an outcome model $f(y\mid x; \theta)$. In situation (a), for example, if the parameter of interest is the mean of the response variable, we can write $U(\theta; x, y)=y-\theta$. Although efficient estimators for the three target parameters are derived in \S \ref{sec:3}, we focus on estimating the parameter for (c) $f(y\mid x; \theta)$ in Setting 2 in this section. Hereafter, the argument on random variables in the functions is abbreviated (e.g., $U(\theta; x,y)=U(\theta)$) when it is obvious. 

\subsection{Weighting estimators}
\label{sec:2.2}

The Horvitz-Thompson method can be employed to develop an unbiased estimator for the total mean of the outcome variable using the sampling weights. 
Using a similar idea, \citet{Binder1983} proposed an estimating equation for $\theta$ with only sampled data:
\begin{align}
\sum_{i=1}^N \delta_i W_i S_{\theta}(X_i, Y_i) = 0, \label{weight}
\end{align}
where $S_\theta (x, y)=\partial \log f(y\mid x; \theta)/\pd \theta$ is the score function for $\theta$.
The estimator obtained from (\ref{weight}) exhibits consistency and asymptotic normality under certain regularity conditions \citep{yuan1998}
because of the unbiasedness of the estimating equation.

\subsection{Conditional maximum likelihood estimator}
\label{sec:2.3}
Let the conditional inclusion probability be denoted as $\pi(x,y)=P(\delta = 1\mid x,y)$.  Using $\pi(x,y)$, \citet{Pfeffermann1999} derived the outcome distribution of sampled units through that of population units:
\begin{align}
f_s(y\mid x; \theta) = f(y\mid x, \delta = 1; \theta) = \frac{ f(y\mid x; \theta)\pi(x, y)}{\int f(y\mid x; \theta)\pi(x, y)\mathrm{d}y}. \label{bayes}
\end{align}
Regarding optimal estimation, Theorem 2.2 in  \citet{godambe1976} showed that an estimating equation based on the score function provides the optimal estimator. Thus, by combining these results, an efficient estimator can be derived as the unique solution to
\begin{align}
\sum_{i=1}^N \delta_i \frac{\pd \log f_s(Y_i\mid X_i; \theta) }{\pd \theta} = \sum_{i=1}^N \delta_i \llp S_\theta(X_i, Y_i) - \frac{E\{ S_\theta(X_i, Y)\pi(X_i, Y)\mid X_i;\theta\}}{E\{ \pi(X_i, Y)\mid X_i;\theta\} }\rrp=0, \label{CML}
\end{align}
where $E\{g(x, Y)\mid x; \theta\}$ is the conditional expectation on $g(x, y)$ with a conditional density function $f(y\mid x;\theta)$.

Generally, $\pi(x,y)$ is unknown and has to be estimated. Suppose that $\pi(x,y)$ is known up to a finite-dimensional parameter. \citet{Pfeffermann09} provided the relationship:
$\pi(x,y) = 1/E_s(W\mid x, y)$, where $E_s(W\mid x, y)=E(W\mid x, y, \delta = 1)$. This relationship implies that the conditional inclusion probability $\pi(x, y)$ can be estimated by regression of weights on the data with sampled units. Therefore, substituting an estimated response mechanism into \eqref{CML}, an ``efficient" estimator for $\theta$ can be obtained. However, to compute the expectation $E_s(W\mid x, y)$, correct model specification for the conditional distribution is imperative, and incorrect model specification can produce a biased estimator.

\section{Semiparametric efficiency bound}
\label{sec:3}
The semiparametric efficiency bound is derived for estimators of $\theta$, which is the smallest asymptotic variance of estimators in a class of regular and asymptotically linear estimators \citep{bickel1998, tsiatis2006}. A key idea in deriving the lower bound under informative sampling is to regard the survey weight as a random variable. Although sampling weights are fixed values from the perspective of the sampling designer, this notion does not hold for data analysts. Data analysts may not have access to all the design variables, rendering the relationship between covariates and sampling weights random from data analysts'  viewpoint.
Then, it is expected that incorporating the information of $W_i$ into the model yields a more efficient estimator. This concept is similar to use the propensity score to obtain more efficient estimators in the literature of missing data and causal inference \citep{robins94, tan2010}. However, to our knowledge, no studies have considered the survey weight as a random variable and derived the semiparametric efficient estimator.

Let each $f(w\mid x, y; \eta_1)$, $f(y\mid x; \eta_2, \theta)$, and $f(x; \eta_3)$ be the density function of $[w\mid (x,y)]$, $[y\mid x]$, and $[x]$, respectively, where $\eta_1$, $\eta_2$, and $\eta_3$ are infinite-dimensional nuisance parameters. The model for complete data is
\begin{align}
 f(x,y,w; \theta, \eta_1, \eta_2, \eta_3)
 = f(w\mid x,y; \eta_1)f(y\mid x; \theta, \eta_2)f(x; \eta_3), \label{full_model}
\end{align}
while the model for observed data under the Setting 1 is
\begin{align}
&\llp P(\delta=1\mid x,y,w)f(x,y,w; \theta, \eta_1, \eta_2, \eta_3)\rrp^\delta \nonumber\\
&\quad \times \llp \int\int\int P(\delta=0\mid x,y,w)f(x,y,w; \theta, \eta_1, \eta_2, \eta_3)\mathrm{d}y\mathrm{d}w\rrp^{1-\delta}, \label{obs_model}
\end{align}
where the sampling mechanism $P(\delta=1\mid x,y,w)=w^{-1}$ is known. If our interest is estimating a mean regression $E(Y\mid x;\theta)=\mu(x;\theta)$,
$\eta_j\;(j=1, 2, 3)$ are infinite-dimensional nuisance parameters that satisfy
\begin{align*}
 \int yf(y\mid x;\eta_2, \theta)\mathrm{d}y = \mu(x;\theta),
\end{align*}
and
\begin{align*}
\int \int f(w\mid x, y;\eta_1)\mathrm{d}w\mathrm{d}z =\int f(y\mid x;\eta_2, \theta)\mathrm{d}y = \int f(x;\eta_3)\mathrm{d}x = 1.
\end{align*}
The model for observed data under Setting 2 is obtained similarly by integrating the function for $\delta = 0$ in \eqref{obs_model} with respect to $x$.

Then, the efficiency bounds for $\theta$ are derived that are not affected by the nuisance parameters in Settings 1 and 2. The key idea is to determine the orthogonal projection of the estimating function on the tangent space $\Lambda$ of the nuisance parameters.  Let $q$ be the dimension of the parameter $\theta$ of interest and $\mathcal{H}$ be a Hilbert space that comprises $q$-dimensional measurable functions $h=h(\delta, X, Y, W)$ such that $E(h)=0$ and $\langle h, h\rangle <\infty$, where the inner product is defined by $\langle h, l\rangle := E(h^\top l)$ for any $h,l\in\mathcal{H}$. Furthermore, let each $\Lambda^{sp,\perp}$ and $\Lambda^\perp$ be the orthogonal nuisance tangent space for the superpopulation model provided in \eqref{full_model} and the model for the sampled data provided in \eqref{obs_model}, respectively. The projection onto a subspace $\tilde{\mathcal{H}} \subset \mathcal{H}$ is denoted by $\Pi(\cdot\mid \tilde{\mathcal{H}})$. Let the efficient score in the superpopulation model be $S^{sp}_{\eff}=\prod(S_{\theta}\mid \Lambda^{sp,\perp})$ and that in the observed model, which is of interest here, be $S_{\eff}=\prod(S_{\theta}\mid \Lambda^\perp)$. 

We derive the efficiency bound of the RAL estimators of $\theta$ under informative sampling by considering the above semiparametric models, in which we  emphasize again  that survey weights are considered as random variables, as non-MAR semiparametric models with the known missing data mechanism. For non-MAR semiparametric models,  \citet{robins94}  and \citet{rotnitzky97}  derived the semiparametric efficiency bound. The following lemma is a modified version of Propositions A1 and A2 in \citet{rotnitzky97}, tailored by treating the sampling weights as random variables and adapting to the framework of informative sampling.

\begin{lem}
\label{lem.a1}
The efficient score is expressed as
\begin{align}
    S_{\eff}(X, Y, W)=\delta W D^*_{\eff} + (1-\delta W)\frac{E\{(W-1)D^*_{\eff}\mid X\}}{E(W-1\mid X)}, \label{eff_score}
\end{align}
where $D^*_{\eff}\in\Lambda^{sp,\perp}$ is the unique solution to
\begin{align}
    \prod\lp WD^*_{\eff}-(W-1)\frac{E\{(W-1)D^*_{\eff}\mid X\}}{E(W-1\mid X)}\;\bigg|\; \Lambda^{sp, \perp}\rp=S^{sp}_{\eff}. \label{eff_eq}
\end{align}
For units with $W=1$, the second term in \eqref{eff_score} is set to be zero and $D^*_{\mathrm{eff}}=S^{sp}_{\mathrm{eff}}$.
\end{lem}

The first term in equation \eqref{eff_score} is essentially the same as the weighting estimators provided in \eqref{weight}, and the second term is an augmented term with zero mean. The augmented term achieves efficiency without sacrificing the consistency of the estimator \citep{robins94}. In the case of informative sampling, the first term is unbiased. Thus, it is evident that the estimator based on the above efficient score has consistency similar to that of weighting estimators.

By solving equation \eqref{eff_eq} for each target parameter, the efficiency bound of the parameter can be obtained. To solve the equation \eqref{eff_eq}, it is sufficient to determine $D^*_{\eff} \in\Lambda^{sp,\perp}$ that fulfills, for any $h\in \Lambda^{sp,\perp}$,
$$\left\langle WD^*_{\eff}-(W-1)\frac{E\{(W-1)D^*_{\eff}\mid X\}}{E(W-1\mid X)} - S^{sp}_\eff, ~h \right\rangle=0.$$
The efficient score $S^{sp}_{\eff}$, the nuisance tangent space $\Lambda^{sp}$, and its orthogonal space $\Lambda^{sp,\perp}$ in the super population model are well known for various target parameters including the three target parameters (a)--(c) discussed in \S \ref{sec:2.1}. See Appendix \ref{Appendix_B} for more details.

However, in general, the solution to \eqref{eff_eq} does not have an explicit form and requires a certain recursive approximation \citep{rotnitzky97}. Fortunately, in the context of informative sampling, the explicit form of $D^*_{\eff}$ can be obtained. 
 First, consider a theory under Setting 1, where covariates $x_i\,(i=1,\dots,N)$ are available throughout the finite population. 
 
\begin{thm}
\label{thm.1}
Consider semiparametric estimators for each parameter (a)--(c) expressed in \S \ref{sec:2} under Setting 1. Then, the efficient scores for the three estimators are
\begin{align}
S_\eff = \delta W D^*_\eff(\theta; X, Y) + (1-\delta W)C^*_\eff(\theta; X), \label{eff_est_eq1}
\end{align}
where $D^*_\eff$ and $C^*_\eff$ are different for the three parameters as follows:\\
\noindent
For $\theta$ defined through $E\{U(\theta; X, Y)\}=0$.
    \begin{align}
    D^*_\eff= U(\theta),\quad C^*_\eff = \frac{E\{(W-1)U(\theta)\mid x\}}{E(W-1\mid x)}. \label{mean_eff}
    \end{align}
\noindent 
For $\theta$ in $\mu(x;\theta)=E(Y\mid x)$.
    \begin{align}
    \begin{split}
        D^*_\eff &= A^*_\eff (x)\{Y-\mu(x;\theta)\},\\
        C^*_\eff &= \lllp E(W-1\mid x) - \frac{\{E(W\varepsilon \mid x)\}^2}{E(W\varepsilon^2\mid x)} \rrrp^{-1} \frac{E(W\varepsilon \mid x)}{E(W\varepsilon^2\mid x)} \dot{\mu}(x;\theta),\label{reg_eff}
    \end{split}
    \end{align}
        where
        $$A^*_\eff(x)=\frac{1}{E(W\varepsilon^2\mid x)}\lllp E(W\varepsilon\mid x)C^*_\eff(x)+\dot{\mu}(x; \theta)\rrrp.$$
\noindent
For $\theta$ in $f(y\mid x; \theta)$.
    \begin{align*}
      D^*_\eff &= \bar{\pi}\llp S_\theta-\frac{E(\bar{\pi} S_\theta\mid x;\theta)}{E(\bar{\pi}\mid x;\theta)}\rrp +\lp 1 - \frac{\bar{\pi}}{E(\bar{\pi}\mid x;\theta)}\rp C^*_\eff(x),\\
      C^*_{\eff}(x) &= \frac{E(\bar{\pi} S_\theta\mid x;\theta)}{E(\bar{\pi}\mid x;\theta)-1},
    \end{align*}
    where 
    $\bar{\pi} (x,y)=\{E(W\mid x,y)\}^{-1}$. 
The lower bound of the asymptotic variance is $\{E(S_{\eff}^{\otimes 2})\}^{-1}$, where $S_{\eff}^{\otimes 2}=S_\eff S_\eff^\top.$
\end{thm}


A similar argument follows in Setting 2, when $N$ is unknown,  and when the sampling mechanism is noninformative. Especially when either $N$ is unknown or the sampling mechanism is non-informative, $C^*_{\eff}=0$ for all the parameters.  See Appendix \ref{Appendix_D} for detailed expressions of efficient scores.

The mathematical formulations of $D^*_{\eff}$ and $C^*_{\eff}$ demonstrate that the expectation of the scores is always zero, regardless of the correctness of the working models. For example, for the parameter of an outcome model, irrespective of the specific form of $\bar{\pi}$ function, it follows that:
\begin{align*}
E(S_\eff)
&=E(D^*_\eff)\\
&=E(\bar{\pi}S_\theta) - E\llp E(\bar{\pi}\mid x;\theta)\frac{E(\bar{\pi}S_\theta\mid x;\theta)}{E(\bar{\pi}\mid x;\theta)}\rrp + E\llp \lp 1- \frac{E(\bar{\pi}\mid x;\theta)}{E(\bar{\pi}\mid x;\theta)}\rp C^*_{\eff}(x)\rrp\\
&=0.
\end{align*}

\begin{rem}
\label{rem.2}
For the parameter (b) regression model in Setting 2, the estimator obtained by fixing $C^*_{\eff}$ to be zero is exactly the same as that proposed in \citet{kim13} when the regression model is linear. Therefore, the estimator proposed by \citet{kim13} is optimal for linear regression when $N$ is unknown, and the estimator in this study can be more efficient when $N$ is known.
\end{rem}









\section{Adaptive estimation}
\label{sec:4}
The efficient score function involves two types of unknown functions: (i) $E\{h(x, Y,W)\mid x\}$ for a specific integrable function $h(x,y,w)$ such as $W\varepsilon$ and $W\varepsilon^2$; (ii) $E(W\mid x,y)$. The first function is estimated by using a working model on $f(y\mid x; \xi)$. The solution to \eqref{weight} with respect to $\xi$, say $\hat{\xi}_{\HT}$, has consistency and asymptotic normality. {Although an optimal estimator can be obtained for $\xi$ instead of $\hat{\xi}_{\HT}$ using the efficient score for (c), the optimal working models are not pursued because it does not affect efficiency of the target parameter.} 
Estimation of the second function requires a model and an estimation method for $E(W\mid x,y)$. This study proposes a reasonable parametric model using beta regression models \citep{ferrari2004}.

\subsection{Parametric working model}
\label{sec:4.1}
Because the sampling weights satisfy $w_i >1\;(i=1, \ldots, n)$, it is assumed that $w_i^{-1}$ is modeled as a Beta distribution $\mathrm{Beta}(m(x_i,y_i) \phi(x_i,y_i),~\{1-m(x_i,y_i)\}\phi (x_i,y_i))$. Thus, the density function is expressed as 
\begin{align*} 
    f(w^{-1} \mid x,y) \propto (w^{-1})^{m \phi -1}(1-w^{-1})^{(1-m)\phi-1},
\end{align*}
and the first and the second moments are
\begin{align*} 
   E(W^{-1}\mid x,y) = m(x,y),\quad
    V(W^{-1}\mid x,y) = \frac{m(x,y)\{1-m(x,y)\}}{1+\phi(x,y)}.
\end{align*}
An example of the mean function is the logistic model:
\begin{equation} 
m(x,y; \beta)=\frac{\exp(\beta_0+\beta_1 x+\beta_2 y)}{1+\exp(\beta_0+\beta_1 x+\beta_2 y)}.
\label{logistic}
\end{equation} 
This is essentially a beta regression model. More details on beta regression can be found in \citet{ferrari2004}.
We can use any function for $\phi(x,y;\gamma)$, but for clarity and conciseness, we choose to use a constant $\phi(x,y)=\phi$ in \S 5 and \S 6.

Unfortunately, the beta regression approach cannot be applied directly because the regression model does not necessarily hold in the sample owing to informative sampling. To avoid this problem, the distribution of sampled data is derived. Recall that if $X$ follows $\mathrm{Beta}(\alpha, \beta)$ then $(1-X)/X$ follows a beta prime distribution $\mathrm{Beta'}(\beta, \alpha)$. Therefore, denoting $O=(1-W^{-1})/W^{-1}=W-1$, $O$ follows $\mathrm{Beta'}(\{1-m(x_i, y_i)\}\phi(x_i,y_i),\; m(x_{i},y_{i})\phi(x_i,y_i))$, and the density function is expressed as
\begin{align*}
    f(o\mid x,y) \propto o^{(1-m) \phi -1}(1+o)^{-\phi}.
\end{align*}
Based on Bayes' theorem and $W^{-1}=(1+O)^{-1}$, the sampled distribution of $O$ is
\begin{align}
    f_s(o\mid x,y) &\propto f(o\mid x,y)P(\delta=1\mid x, y, w)= o^{(1-m) \phi -1}(1+o)^{-\phi-1}, \label{odds_obs}
\end{align}
which implies $O\mid (x,y,\delta=1) \sim \mathrm{Beta'}(\{1-m(x,y)\}\phi(x,y),~m(x,y)\phi(x,y) +1)$. Upon modeling $m(x,y; \beta)$ and then estimating $\beta$ and $\gamma$, two conditional distributions are obtained using the fact that $X$ follows $\mathrm{Beta}'(a,b)\,(a>0,b>1)$ implies $E(X) = a/(b-1)$, we have
\begin{align}
  E(W\mid x, y) &= 1+E(O\mid x,y) = \frac{\phi(x,y;\gamma)-1}{m(x,y;\beta)\phi(x,y;\gamma)-1},\label{E_W}\\
  E_s(W\mid x, y) &= 1+E_s(O\mid x,y) = \frac{1}{m(x,y;\beta)}, \label{E1_W}
\end{align}
where \eqref{E_W} holds for all $x,y$ that satisfy $m(x,y;\beta)\phi(x,y;\gamma)>1$. To estimate $\beta$ and $\gamma$, the maximum likelihood estimation based on the distribution \eqref{odds_obs} is used. Using the estimated $\hat{\beta}$ and $\hat{\gamma}$, an estimator of $\bar{\pi}(x,y) = 1/E(W\mid x, y)$ can be obtained based on \eqref{E_W}. However, there is a possibility that denominator in \eqref{E_W} being to negative when the true value of $\phi(x,y)$ is small. One ad-hoc remedy for the problem is to estimate $\beta$ by the ordinary least squares with \eqref{E1_W}, and subsequently estimate $\gamma$ by the methods of moment.

   Identifiability issues are often challenging in missing not at random (MNAR) situations. However, in our setup, known sampling weights mitigate this problem. We can show the identifiability of the beta regression model, meaning that $f_s(o\mid x,y; \beta, \gamma)=f_s(o\mid x,y; \beta', \gamma')$ a.s. implies $\beta=\beta'$ and $\gamma=\gamma'$, under a mild condition. Specifically, from the equality $f_s(o\mid x,y; \beta, \gamma)=f_s(o\mid x,y; \beta', \gamma')$ a.s. and condition \eqref{E1_W}, which represents the equality of conditional means, it follows that $m(x,y;\beta)=m(x,y;\beta')$. Similarly, by applying the equality on conditional variances, we find $\phi(x,y;\gamma)=\phi(x,y;\gamma')$. Therefore, the identifiability of $m(x,y;\beta)$ and $\phi(x,y;\gamma)$, as specified by condition (B2) in Appendix B, ensures that $\beta=\beta'$ and $\gamma=\gamma'$.

\begin{rem}
The choice of adopting beta regression modeling is motivated by the fact that both \( f(w \mid x, y) \) and \( f_s(w \mid x, y) \) are known parametric distributions. We can utilize other positive continuous distributions for weights $W$.
Indeed, any \( f_s(w \mid x, y) = f(w \mid x, y, \delta = 1) \) model yields
\begin{equation*}
    E(W \mid x, y) = \frac{E_s(W^2 \mid x, y)  }{E_s(W \mid x, y) }.
\end{equation*}
However, estimating both $E_s(W^2\mid x,y)$ and $E_s(W\mid x,y)$ and computing these expectations for all \( (x_i, y_i), \) \( (i = 1, \dots, N) \) are practically demanding.
\end{rem}

If the sampling mechanism is stratified sampling, it is reasonable to assume that $w^{-1}$ follows a beta distribution in each stratum. Let $G$ be an indicator that takes $G=g\;(g=1, \ldots, H)$ when the sample belongs to $g$. The joint model specification of $(w, g, y)$ given $x$ can be decomposed as 
$$ [w, g, y\mid x] = [w | x, y, g] \cdot [ g \mid x, y] \cdot [y \mid x], $$
where 
\begin{align}
\begin{split}
 W^{-1} \mid (x, y, G=g) &\sim \mathrm{Beta}(m_g(x, y; \beta_g)\phi_g, \{1-m_g(x,y; \beta_g)\}\phi_g), \\ 
 G \mid x, y &\sim \mathrm{Multi}(1; (p_1, \cdots, p_H) ), \\
p_g = P(G=g \mid x, y) &= \frac{ \exp (\alpha_{g0} + \alpha_{g1} x+ \alpha_{g2} y)}{ \sum_{h=1}^H 
 \exp (\alpha_{h0} + \alpha_{h1} x+ \alpha_{h2} y)
 }, 
 \end{split}\label{mix_model}
 \end{align} 
with $\alpha_{g0}=\alpha_{g1} = \alpha_{g2}=0$ for $g=1$. Using the above joint distribution, the sampled likelihood can be derived to obtain an efficient estimator. The detailed scheme for the EM algorithm is provided in Appendix \ref{Appendix_C}.
As shown in \eqref{E_W}, $E(W\mid x,y)$ can be calculated using the estimated parameters as
\begin{align}
E(W\mid x,y; \hat{\alpha}, \hat{\beta}, \hat{\phi}) = \sum_{h=1}^H p_h(\hat{\alpha}_h) \frac{\hat{\phi}_h - 1}{m_h(x, y; \hat{\beta}_h)\hat{\phi}_h - 1}. \label{mix_barpi}
\end{align}

Given that $w^{-1} \mid (x, y, G=g) = w^{-1} \mid (G=g)$, the weight mechanism is 
\begin{equation} 
W^{-1} \mid (x, y, G=g) \sim \mathrm{Beta}( m_g\phi_g, (1-m_g)\phi_g).
\label{no_depend}
\end{equation} 
This weight model is used in the real data analysis in \S\ref{sec:6}.

Next, we investigate the large sample property of the proposed estimator in Setting 1. The property in Setting 2 is obtained in a similar way.
Let the parameter vector in the weight model be $\psi = (\phi^\top, \alpha^\top, \beta^\top{,\gamma^\top})^\top$. After obtaining the working models, an estimating equation for $\theta$ based on the efficient score can be expressed as a function of $\hat{\xi}_{\HT}$ and $\hat{\psi}$
\begin{align}
\sum_{i=1}^N S_{\eff}(X_i, Y_i, Z_i, W_i; \theta, \hat{\xi}_{\HT}, \hat{\psi})=0, \label{eff_adapt_eq}
\end{align}
where $\hat{\xi}_{\HT}$ is the estimated parameter in the working model $f(y\mid x; {\xi})$ and 
 $\hat{\alpha}$ is used only when the sampling mechanism is a stratified sampling. Before discussing the theoretical properties of the proposed estimator, consider two examples of adaptive estimators.
 
 \begin{example}[Optimal estimation for mean]
 \label{ex:1}
Consider the estimation of $\theta= E(Y)$ in Settings 1 and 2. When $N$ is unknown, Corollary \ref{cor.2} implies that the Horvitz-Thompson estimator is the best. However, when $N$ is known in Setting 1, as per \eqref{mean_eff}, the optimal estimator is 
\begin{align}
\hat{\theta} = \frac{1}{N} \sum_{i=1}^N\llp \delta_i W_i Y_i + (1-\delta_i W_i) \frac{\int y\{E(W\mid X_i,y; \hat{\psi})-1\}f(y\mid X_i; \hat{\xi}_{\HT})dy}{\int E(W\mid X_i,y; \hat{\psi})-1\}f(y\mid X_i; \hat{\xi}_{\HT})dy} \rrp, \label{mean_eff1_sample}
\end{align}
where $E(W\mid x,y; \hat{\psi})$ is defined in \eqref{mix_barpi}. The augmented term can also be estimated using weighted regression. Suppose that we have a model for \(g(x;\zeta) = E\{Y(W-1) \mid x\} / E(W-1 \mid x)\), such as a B-spline expansion. The parameter in the model can be estimated by directly minimizing
\[
\sum_{i=1}^n \delta_i W_i (W_i-1)\{Y_i-g(X_i;\zeta)\}^2,
\]
which is more convenient for this specific parameter.

In Setting 2, as per \eqref{mean_eff2}, the estimator is 
\begin{align*}
\hat{\theta} = \frac{1}{N} \sum_{i=1}^N\llp \delta_i W_i Y_i + (1-\delta_i W_i) \frac{E\{Y(W-1)\}}{E(W-1)} \rrp.
\end{align*}
Based on Bayes' theorem, similar to \eqref{bayes}, the second term can be expressed as
\begin{align*}
\frac{E\{Y(W-1)\}}{E(W-1)} = \frac{E_s\{YW(W-1)\}}{E_s\{W(W-1)\}}.
\end{align*}
Therefore, the optimal estimator is obtained without using any working model as
\begin{align}
\hat{\theta} = \frac{1}{N} \sum_{i=1}^N\llp \delta_i W_i Y_i + (1-\delta_i W_i) \frac{\sum_{j=1}^N \delta_j Y_jW_j (W_j-1)}{\sum_{j=1}^N \delta_j W_j (W_j-1)}\rrp.
\label{mean_eff2_sample}
\end{align}
This estimator is more efficient than the Hajek estimator, $\sum_{i=1}^n W_i Y_i / \sum_{i=1}^n W_i$, in Setting 2, due to the incorporation of information about $N$.

\end{example}

\begin{example}[Optimal estimation for regression models]
\label{ex:2}
Consider the estimation of regression models $\mu(X; \theta)$. When $N$ is unknown, Corollary \ref{cor.2} implies that the optimal estimating equation can be expressed as
\begin{align*}
    \sum_{i=1}^N \delta_i W_i \frac{\dot{\mu}(X_i; \theta)}{\int E(W\mid X_i, y; \hat{\psi})\{y - \mu(X_i; \theta)\}^2 f(y\mid X_i; \hat{\xi}_{\HT})\mathrm{d}y}\{Y_i - \mu(X_i; \theta)\} = 0.
\end{align*}
As stated in Remark \ref{rem.2}, this estimating equation is the same as that proposed in \citet{kim13} when the regression model is linear. However, a more efficient estimator can be constructed by using additional information with $C^*_{\eff}$ and $\tilde{C}^
*_{\eff}$ in \eqref{reg_eff} and \eqref{reg_eff2}, which can be estimated in a manner similar to that of Example \ref{ex:1}.
\end{example}
 


Our estimator balances the efficiency of estimation against the accuracy of the assumed model. Consequently, incorrect specifications of these models can lead to significantly inefficient results. Nevertheless, we posit that the cost associated with estimating unknown functions remains low, given that the precision of these models can be indirectly assessed by comparing the estimated asymptotic variances of both the Horvitz-Thompson and our suggested estimators. Should the models be substantially incorrect, the estimated asymptotic variance will exceed that of the Horvitz-Thompson estimator, indicating a need to either revisit the model assumptions or consider adopting the Horvitz-Thompson estimator as a more reliable alternative. For detailed information on the numerical effects of model misspecification, refer to Scenario S4 in \S 5.

Let the unique solution of \eqref{eff_adapt_eq} be $\hat{\theta}$. We relegated the regularity conditions B1--B6 to Appendix \ref{Appendix_A} because the following asymptotic theory is obtained using the standard argument on the methods of moment estimator. The proposed estimator exhibits consistency even if all the working models are wrong,  and attains the semiparametric efficiency bound if all the working models are correct.

\begin{thm}
\label{thm.2}
Under Setting 1 and regularity conditions B1--B6 in Appendix \ref{Appendix_A}, the estimator $\hat{\theta}$ shows consistency and $\sqrt{N}( \hat{\theta} - \theta )$ weakly converges to the normal distribution with mean zero and variance
\begin{align}
    \lllp E\left\{\frac{\partial S^{*}_{\mathrm{eff}}(\theta)}{\partial \theta^\top}\right\}\rrrp ^{-1} E[\{S^{*}_{\mathrm{eff}}(\theta)\}^{\otimes 2}]\lllp E\left\{\frac{\partial S^{*}_{\mathrm{eff}}(\theta)}{\partial \theta^\top}\right\}\rrrp^{-1} \label{sd_formula}
\end{align}
as $N\to\infty$, where $S^*_{\mathrm{eff}}(\theta)=S_{\mathrm{eff}}(\theta, \xi^*_{\HT}, \psi^*)$ , $\xi^*_{\HT}$ and $\psi^*$ are the probability limits of $\hat{\xi}_{\HT}$ and $\hat{\psi}$, respectively. Furthermore, if all working models are correct, the asymptotic variance is equal to $\{E(S^{\otimes 2}_{\mathrm{eff}}(\theta))\}^{-1}$, which is the semiparametric efficiency bound derived in \S \ref{sec:3}.
\end{thm}

\subsection{Nonparametric working model}
\label{sec:4.2}
We considered a reasonable parametric model on $E(W\mid x,y)$ or $\bar{\pi}(x,y)$ in \S \ref{sec:4.1}. 
Here, we propose two types of fully nonparametric working models with the kernel ridge estimation.
Let $k(\cdot, \cdot): \mathbb{R}^{1+d_X}\times\mathbb{R}^{1+d_X}$ $\to$ $\mathbb{R}$ 
be a positive-definite kernel with the reproducing kernel Hilbert space (RKHS) $\Hk$, 
where $d_X$ is the dimension of $X$.
For simplicity of notation, without loss of generality, we assume that $\delta_i = 1$ for $i=1,\dots,n$.
Write $V_i = (X_i,Y_i)$ and $v_i = (x_i,y_i)$.
We write $K_n$ for the $n\times n$ gram matrix that has $k(v_i, v_j)$ in the $(i,j)$-th component.

We can also consider the weighted kernel ridge regression:
\[
\min_{f\in \Hk} \left[ \sum_{i=1}^N \delta_iW_i \{W_i - f(v_i)\}^2 + \lambda_n \|f\|_{\Hk}^2 \right].
\]
By the representer theorem, the resulting weighted kernel Ridge estimator is represented by
\banum
\frac{1}{\hat{\bar{\pi}}(v)}=\hat{f}_n(v)=\sum_{i=1}^n \hat{a}_i k(v, v_i),
\label{eq:wkrr}
\eanum
where $D_n := \mathrm{diag}(W_1,\dots,W_n)$, and 
\ba
\hat{a} = (\hat{a}_1, \dots, \hat{a}_n)^\top 
&= \argmin_{a\in\mathbb{R}^n}\left[\sum_{i=1}^n W_i (W_i - K_na)^2 + \lambda_n a^\top K_na\right]\\
&= (D_nK_n+\lambda_n I_n)^{-1}D_n J_n.
\ea

Alternatively, since we have $E(W\mid V=v) = E_s(W^2\mid V=v)/E_s(W\mid V=v)$, natural estimators for both $E_s(W\mid v)$ and $E_s(W^2\mid v)$ can be constructed by the following kernel ridge regressions $W^j\,(j=1,2)$ on $V$:
\ba
\min_{f^{(j)}\in \Hk} \left[ \sum_{i=1}^n \{W^j_i - f^{(j)}(v_i)\}^2 + \lambda^{(j)}_n \|f^{(j)}\|_{\Hk}^2 \right], \quad (j=1,2)
\ea
where $\lambda^{(1)}_n$ and $\lambda^{(2)}_n$ are tuning parameters, and $\|\cdot\|_{\Hk}$ be the norm of the RKHS $\Hk$. By the representer theorem, the resulting kernel Ridge estimators are represented by
\ba
\hat{f}^{(j)}_n(v)=\sum_{i=1}^n \hat{a}^{(j)}_i k(v, v_i), \quad (j=1,2)
\ea
where $J_n:=(W_1,\dots,W_n)^\top$,
$\hat{a}^{(1)} = (\hat{a}^{(1)}_1, \dots, \hat{a}^{(1)}_n)^\top = (K_n+\lambda^{(1)}_n I_n)^{-1} J_n$, and
$\hat{a}^{(2)} = (\hat{a}^{(2)}_1, \dots, \hat{a}^{(2)}_n)^\top = (K_n+\lambda^{(2)}_n I_n)^{-1} J_n^2$.
If there exists a constant $C>0$ such that $1<W<C$ almost surely, as we will assume later in condition N1, the following clipped version of $\hat{f}^{(j)}_n(v)$ may provide better results:
\[
\hat{f}_{\mathrm{clip},n}^{(j)}(v) = 
\begin{cases}
 \hat{f}^{(j)}_n(v) & \text{if }\; 1\le \hat{f}^{(j)}_n(v) \le C,\\
C^j & \text{if }\; C^j< \hat{f}^{(j)}_n(v),\\
1 & \text{if }\; \hat{f}^{(j)}_n(v)< 1.
\end{cases}
\]
As described later, this assumption is natural in the informative sampling setup.
To notation, we also use $\hat{f}^{(j)}_n(v)$ for $\hat{f}_{\mathrm{clip},n}^{(j)}(v)$.
Thus, we can estimate $\bar{\pi}(v)$ by 
\begin{align}
\widehat{\bar{\pi}}(v)=\hat{f}^{(1)}_n(v)/\hat{f}^{(2)}_n(v). \label{eq:sp}
\end{align}
Here, we note that the difference between $\widehat{\bar{\pi}}(v)$ and $\bar{\pi}(v)$ can be bounded by 
the differences between $\hat{f}^{(j)}_n(v)$ and $E_s(W^j\mid V=v)\;(j=1,2)$.
That is, we have
\begin{align}
\begin{split}
|\widehat{\bar{\pi}}(v)-\bar{\pi}(v)|
&\le Const.\times \{ |\hat{f}^{(1)}_n(v)-E_s(W\mid V=v)|\\
&\quad + |\hat{f}^{(2)}_n(v)-E_s(W^2\mid V=v)|\}. 
\end{split}\label{bound}
\end{align}

Unfortunately, however, a plug-in estimator with the estimated nonparametric working models \eqref{eq:wkrr} or \eqref{eq:sp} into
\begin{align}
\sum_{i=1}^N S_{\eff}(X_i, Y_i, W_i; \theta, \hat{\bar{\pi}}(\cdot))=0, \label{eff_adapt_eq_semi}
\end{align}
may be biased because the convergence rate of $\hat{\bar{\pi}}(\cdot)$ is generally slower than $n^{-1/2}$. To overcome the difficulty, we use the technique called, double/debiased machine learning (DML), splitting the original dataset $\mathcal{D}$ into two parts $\mathcal{D}_1$ and $\mathcal{D}_2$ \citep{chernozhukov2018double}.
At the first step, construct nonparametric estimators $\hat{\bar{\pi}}_1(\cdot)$ and $\hat{\bar{\pi}}_2(\cdot)$ for $\bar{\pi}(\cdot)$ by using $\mathcal{D}_1$ and $\mathcal{D}_2$, respectively.
Next, 
we solve the following efficient estimating equations for $j=1,2$ and denote the solutions as $\hat{\theta}_j\,(j=1,2)$:
\[
\sum_{i \in \mathcal{D}_j} S_{\eff}(X_i, Y_i, W_i; \theta, \hat{\bar{\pi}}_{3-j}(\cdot)) = 0.
\]
The DML estimator is constructed by calculating the average $\hat{\theta}_{\mathrm{DML}} = (\hat{\theta}_1 + \hat{\theta}_2) / 2$.
We can diminish the randomness inherent in splitting the data sets by performing multiple random splits, say $L$ times. For each $l$-th split dataset, we compute the DML estimators $\hat{\theta}^{(l)}_{\mathrm{DML}}\,(l=1,\dots, L)$, and then take the average or median of these DML estimators. The tuning parameters $\lambda^{(1)}_{n}$, $\lambda^{(2)}_{n}$, and $\lambda_{n}$ can be chosen using generalized cross-validation, which approximates leave-one-out cross-validation and facilitates faster computation (e.g., see \citep[p.244]{ELS2008}).

So far, we have proposed two types of nonparametric working models for $\bar{\pi}(v)$: (i) direct but weighted estimation in \eqref{eq:wkrr}; (ii) separate estimation of $E_s(W^j\mid v),(j=1,2)$ in \eqref{eq:sp}. We anticipate that the differences in the effects produced by these two methods would be insignificant when $n$ is sufficiently large. However, for small sample sizes, our empirical numerical studies showed that the weighted kernel ridge estimator underperformed compared to the separate estimate of $E_s(W^j\mid v),(j=1,2)$, probably due to weighting. Therefore, we will only demonstrate the performance of the DML estimator with the separately estimated working model in the following sections. 

Thanks to the orthogonality of the estimating equation, we obtain the desired large sample property for $\hat{\theta}_{\mathrm{DML}}$ under very mild constraints for the two types of nonparametric working models. 
We assume the following regularity conditions for the working model with weighted kernel ridge regression defined in \eqref{eq:wkrr}:
\begin{itemize}
\item[N1.] There exists $C>0$ such that $1< W <C$ almost surely.
\item[N2.] There exists $\kappa>0$ such that $k(v,v)<\kappa$ for all $v$.
\item[N3(a).] Let the true conditional expectation $E(W|V=v)$ be $f_0(v)$ and $f_0\in \Hk$.
\item[N4(a).] Let $T:\Hk\rightarrow \Hk$ be the integral operator with kernel $k$:
\[
Tf = \int k(\cdot,v)f(v)dP_V(v),\quad f \in \Hk,
\]
where $P_V$ is the probability measure of $V$.
There exist $c \in (1,2]$, $R>0$, and $g\in \Hk$ such that $f_0 = T^{(c-1)/2}g$ and $\|g\|_{\Hk}^2\le R$.
\item[N5(a).]
There exist $b>1$ and $\alpha,\beta>0$ such that $\forall i \ge 1;\;\alpha \le i^b \mu_i \le \beta$, where $\mu_i$ is the $i$-th eigenvalue of $T$.
\end{itemize}
Instead of N3(a)--N5(a),
the following three conditions are required for the working model with separately estimated kernel ridge regressions defined in \eqref{eq:sp} while the first two conditions [N1] and [N2] are the same:
\begin{itemize}
\item[N3(b).] For $j=1,2$, let the true conditional expectation $E_s(W^{j}|V=v)$ be $f_0^{(j)}(v)$ in $\Hk$.
\item[N4(b).] Let $\tilde{T}:\Hk\rightarrow \Hk$ be the integral operator with kernel $k$:
\[
\tilde{T}f = \int k(\cdot,v)f(v)dP_{V_s}(v),\quad f \in \Hk,
\]
where $P_{V_s}$ is the probability measure of $V$ given $\delta=1$.
For $j=1,2$, there exist $c^{(j)} \in (1,2]$, $\tilde{R}>0$, and $g^{(j)}\in \Hk$ such that $f_0^{(j)} = \tilde{T}_s^{(c^{(j)}-1)/2}g^{(j)}$ and $\|g^{(j)}\|_{\Hk}^2\le \tilde{R}$.
\item[N5(b).]
There exist $\tilde{b}>1$ and $\tilde{\alpha},\tilde{\beta}>0$ such that 
$\forall i \ge 1;\;\tilde{\alpha} \le i^{\tilde{b}} \tilde{\mu}_i \le \tilde{\beta}
$, where $\tilde{\mu}_i$ is the $i$-th eigenvalue of $\tilde{T}$.
\end{itemize}

Condition N1 is analogous to the positivity assumption that is frequently assumed in the missing data and causal inference literature. However, under informative sampling, this condition is less stringent than those in typical setups because $W_i\,(i=1,\dots, n)$ are known. Conditions N2, N3(a)--N5(a), and N3(b)--N5(b) are necessary to ensure the subsequent convergence rate of the kernel ridge estimators as detailed in \citet{CaponnettoDeVito07}.

\begin{lem}
\label{lem.a2}
Under conditions N1, N2, and N3(a)--N5(a), we have $\|\hat{f}_n-f_0\|_{L^2(P_V)} = O_P(n^{-bc/(bc+1)}) = o_P(n^{-1/4})$ for the working model with weighted kernel ridge regression, as defined in \eqref{eq:wkrr}. Similarly, under conditions N1, N2, and N3(b)--N5(b), we obtain $\|\hat{f}^{(j)}_n-f^{(j)}_0\|_{L^2(P_{V_s})} = O_P(n^{-\tilde{b}c^{(j)}/(\tilde{b}c^{(j)}+1)}) = o_P(n^{-1/4})$ for $j=1,2$ for the working model with separately estimated kernel ridge regressions, as defined in \eqref{eq:sp}.
\end{lem}

The desired asymptotic properties of the DML estimator can be directly derived from the inequality \eqref{bound}, Lemma \ref{lem.a2}, and Theorem~3.3 of \citet{chernozhukov2018double}.
\begin{thm}
\label{thm.3}
Under Setting 1 and regularity conditions N1, N2, and N3(a)--N5(a), the DML estimator $\hat{\theta}_{\mathrm{DML}}$, equipped with the working model of weighted kernel ridge regression as defined in \eqref{eq:wkrr} and $\lambda_n=O(n^{b/(bc+1)})$, exhibits consistency. Furthermore, $\sqrt{N}( \hat{\theta}_{\mathrm{DML}}-\theta )$ weakly converges to a normal distribution with zero mean and variance $\{E(S^{\otimes 2}_{\mathrm{eff}}(\theta))\}^{-1}$, which corresponds to the semiparametric efficiency bound derived in \S \ref{sec:3}. In the case where the working model utilizes separately estimated kernel ridge regressions as defined in \eqref{eq:sp}, the required convergence rate for the tuning parameter alters to $\lambda^{(j)}_n=O(n^{\tilde{b}/(\tilde{b}c^{(j)}+1)})$ for both $j=1,2$.
\end{thm}



\section{Numerical study}
\label{sec:5}

We conducted numerical studies to demonstrate the consistency of our proposed estimators and to investigate the effects of the correctness of working models on efficiency in Scenarios S1--S3. Additionally, we explore the sensitivity of these estimators against model misspecification of working models in Scenario S4. Finite populations with values $(x_i, y_i, z_i, w_i)\;(i=1, \ldots, N)$ were generated repeatedly 1,000 times. Additionally, the two covariates, response variables, and survey weights were independently generated from $x_i\sim N(0, 1/2)$, $z_i\sim N(0, 1/2)$, $y_i\sim N(x_i - z_i, 1/2)$, and $w_i^{-1} \sim \mathrm{Beta}(m(x_i,y_i,z_i)\phi, \{1-m(x_i,y_i,z_i)\}\phi)$ with $\phi = 2500$. Three scenarios for $m(x,y,z)$ were prepared: S1. $\mathrm{logit}\{m(x,y,z)\}= -3.2$; S2. $\mathrm{logit}\{m(x,y,z)\}=-3.6+ 0.75x +0.5y$; S3.  $\mathrm{logit}\{m(x,y,z)\}=-3.3 + 0.3x + 0.3z + 0.1y^2$. We used Poisson sampling with the inclusion probability $w_i^{-1}$ to select the samples. 
The sampling indicator $\delta_i$ was generated independently from the binomial distribution with probability $w_i^{-1}$. The number of finite populations $N=10,000$, but the size of the sampled units $n$ was approximately 400 in all scenarios.

Based on the above setting, the conditional distribution of $Y\mid x$ is normal: $Y\mid x\sim N(a+b x,~\sigma^2)$, where $\theta = (a, b, \sigma^2)$, and the true value of the target parameter is $\theta = (0, 1, 1)$. 
In all the scenarios, a parametric working model for $w_i$ is given by  $W_i\mid (x_i, y_i) \sim \mathrm{Beta}(m(x_i,y_i) \phi,~\{1-m(x_i,y_i)\}\phi)$, where 
\begin{align}
\mathrm{logit}\{m(x,y)\}=\alpha_0 + \alpha_1 x + \alpha_2 y .  \label{work_res}
\end{align}
 Thus, this model specification is correct for Scenarios S1--S2, but incorrect for S3. 

The target parameter $\theta = (a, b, \sigma^2)$ was estimated using six methods: (i) CC: complete case analysis, that is, the solution to equation \eqref{weight} with $W_i\equiv 1$; (ii) HT: Horvitz-Thompson estimator, that is, the solution to \eqref{weight} with $W_i = w_i$; (iii) CML: conditional maximum likelihood estimation with a parametric working model \eqref{work_res} for $m$ in all scenarios; (iv) $\mathrm{Eff}^{ij}_{\mathrm{reg}}\;(\{i,j\}=\{0,0\}, \{1,0\}, \{1,1\})$: proposed adaptive estimator for the regression model with the same parametric working model $m$ as CML. The superscript ``$i$" indicates whether $N$ is known, and ``$j$" indicates whether $x_i$ is available for units $\delta_i=0$ (i.e., the indicator of whether the Setting is 1 or 2); (v) $\mathrm{Eff}^{ij}_{\mathrm{out}}\;(\{i,j\}=\{0,0\}, \{1,0\}, \{1,1\})$: proposed adaptive estimator for the outcome model with the same parametric working model $m$ as CML. The superscripts $\{i,j\}$ are the same as the above ones; (vi)  $\mathrm{Eff}^{11}_{\mathrm{out},\mathrm{DML}}$: proposed adaptive estimator for the outcome model with the nonparametric working model when $i=1$ and $j=1$. We considered this DML estimator only in Scenario S3 to see the semiparametric efficiency bound in S3. To maintain stability in the results, we designated $L$ as 11, which is the number of random splits discussed in \S \ref{sec:4.2}. This number can be set as large as possible, time permitting. In the subsequent real data analysis in \S \ref{sec:6}, we used $L=1000$. 
    For the DML estimator, we use the radial basis function  $k(x,y) = \exp(-\|x-y\|^2/\nu^2)\;(\nu>0)$ as the kernel function. 
    We set the parameter $\nu$ to be half of the median of pairwise squared distances $\|v_i-v_j\|^2\;(i,j=1,\dots,n;\;i\neq j)$, which is the so-called median heuristic.
    Using generalized cross-validation, we chose the tuning parameters $\lambda^{(1)}_n$ and $\lambda^{(2)}_n$ from the candidate values ranging from 0.01 to 1.00, with an interval of 0.01.

\begin{figure}[htbp]
		 \begin{center}
		  \includegraphics[width=140mm]{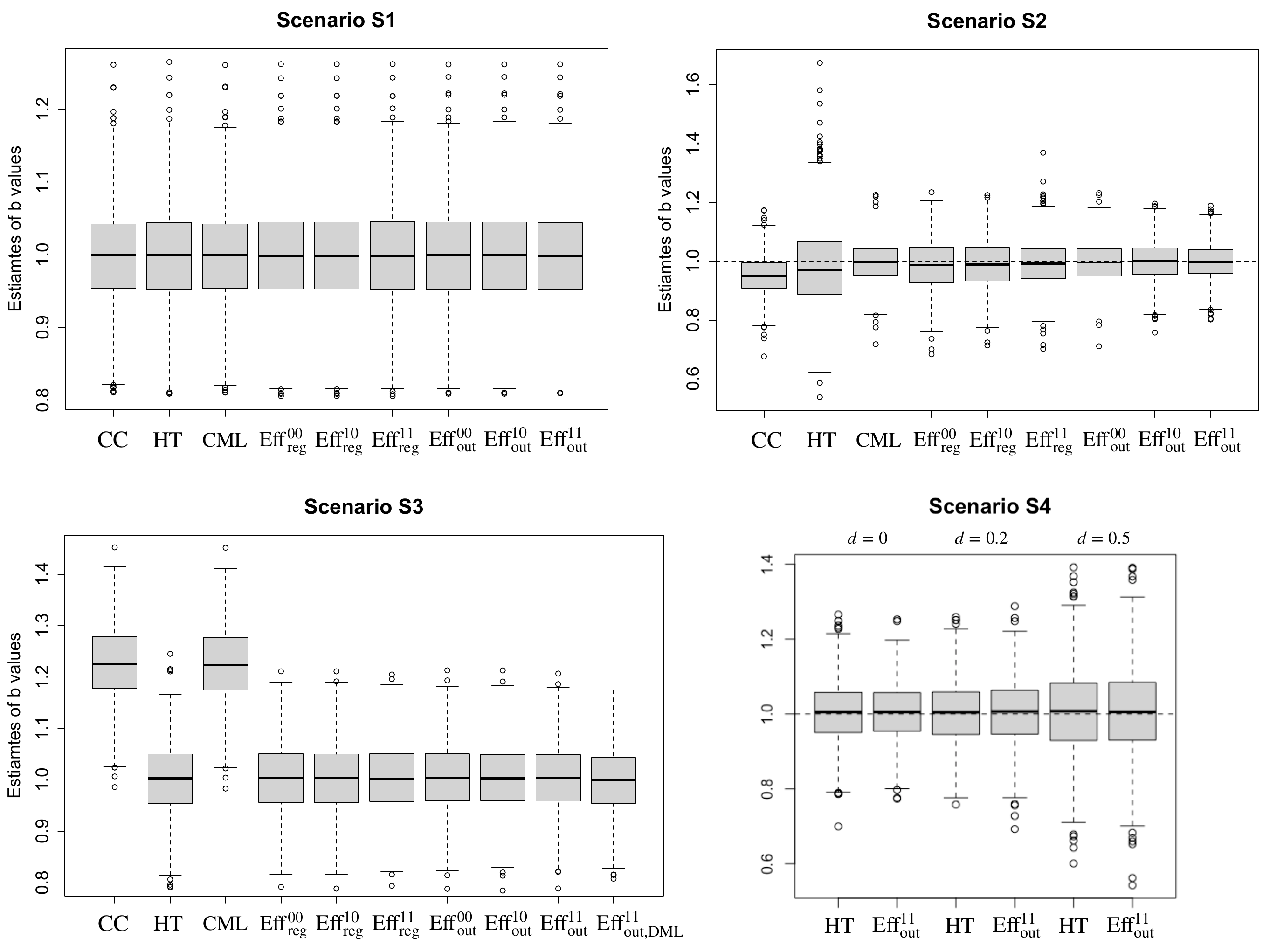}
		 \end{center}
		 \caption{Boxplot for $\hat{b}$ in Scenarios S1--S4: CC (complete case analysis), HT (Horvitz-Thompson type estimator), CML (conditional maximum likelihood estimator), $\mathrm{Eff}_{\mathrm{reg}}^{ij}$ (proposed estimator for the regression parameter), $\mathrm{Eff}_{\mathrm{out}}^{ij}$ (proposed estimator for the outcome model) for $\{i,j\}=\{0,0\}, \{1,0\}, \{1,1\}$, and  $\mathrm{Eff}_{\mathrm{out,DML}}^{11}$ (proposed estimator for the outcome model with the nonparametric working model)  The superscript ``$i$" indicates whether $N$ is known, and ``$j$" indicates whether information of $x$ for units $\delta=0$ is available. The bottom-right figure shows the result in Scenario S4 when $d=0$, $0.2$ and $0.5$.}
    \label{fig:2}
\end{figure}

Figure \ref{fig:2} presents boxplots of estimates for the slope parameter $b$. Under Scenario S1, the sampling weights were equal and no additional information could be inferred from these weights. Consequently, the performance of the estimators was roughly equivalent. Under Scenario 2, the CC estimator exhibited bias due to the sampling weights, and the CML estimator worked well as the sampling weights were correct. Moreover, the semiparametric efficient estimators that employed parametric models were efficient, with efficiency improving as information increased ($E^{00} \rightarrow E^{10} \rightarrow E^{11}$). Under Scenario 3, misspecified weight models caused bias for the CML estimators. However, the proposed estimators with the parametric working model demonstrated good performance due to the robustness against misspecification of the working model as discussed in Theorem \ref{thm.2}. Additionally, the DML estimator slightly outperformed the other estimators because it theoretically attains the semiparametric efficiency bound. 
Table \ref{tab:2} outlines the 95\% coverage probabilities with estimated standard errors using the \eqref{sd_formula}. The coverage probabilities were acceptable, closely aligning with the nominal level across all scenarios.

\begin{table}[hbtp]
  \caption{95\% coverage probability of estimates for $b$ with our 6 proposed estimators. }
  \label{tab:2}
  \centering
  \begin{tabular}{cccccccc}
    \hline \hline
                         Scenario &   $\mathrm{Eff}^{00}_{\mathrm{reg}}$ & $\mathrm{Eff}^{10}_{\mathrm{reg}}$ &  $\mathrm{Eff}^{11}_{\mathrm{reg}}$ &    $\mathrm{Eff}^{00}_{\mathrm{out}}$ & $\mathrm{Eff}^{10}_{\mathrm{out}}$ & $\mathrm{Eff}^{11}_{\mathrm{out}}$ \\ \hline
     S1                 & 0.946 & 0.947 & 0.943 & 0.951 & 0.951 & 0.949 \\
     S2                 & 0.928 & 0.922 & 0.942 & 0.953 & 0.927 & 0.948 \\ 
     S3                 & 0.941 & 0.934 & 0.933 & 0.938 & 0.933 & 0.934 \\ 
    \hline \hline
  \end{tabular}
\end{table}

Next, we assessed the robustness of our method to model misspecification. We repeatedly generated finite populations with values \((x_i, y_i, v_i); (i=1, \ldots, N)\) 1,000 times. Each \(x_i\), \(v_i\), and \(e_i\) independently follows a \(N(0,1)\) distribution, and \(y_i\) is determined by \(y_i = a + b x_i + c v_i + d x_i^2 + e_i\), with coefficients \((a, b, c) = (0,1,1)\), examining \(d\) under three conditions: \(d = 0\), \(0.2\), or \(0.5\). In Scenario S4, we implemented probability proportional to size sampling with replacement, setting the sample size to \(n=400\). This is proportional to \(m_i = \mathrm{Round}(500 \times \mathrm{expit}(1+v_i))\), resulting in sampling weights of \(w_i = \sum_{i=1}^n m_i / (n m_i)\), where $\mathrm{Round}(x)$ represents the rounded value of $x$.

We modeled the weights using a beta regression model with the mean structure:
$$
m(x,y;\beta)=\frac{\exp(\beta_0+\beta_1 x + \beta_2 y + \beta_3x^2+ \beta_4 xy + \beta_5 y^2)}{1+\exp(\beta_0+\beta_1 x + \beta_2 y + \beta_3x^2+ \beta_4 xy + \beta_5 y^2)},
$$
alongside a constant dispersion parameter. The model intended for comparison, as in Scenarios S1--S3, posits \(Y\mid (X=x) \sim N(a+bx, \sigma^2)\). Thus, our applied model is misspecified with the true model structure, highlighting that the target model is also misspecified unless \(\beta_3 \neq 0\). 
The bottom right in Figure \ref{fig:2} illustrates that both the Horvitz-Thompson estimator and our proposed estimator share identical means yet differ in efficiency. When the target model is accurately specified, the proposed adaptive estimator demonstrates greater efficiency. However, as the degree of model misspecification increases, the asymptotic variance of the estimator also enlarges, which is inherent to semiparametric estimators.


\section{Real data analysis}
\label{sec:6}

\begin{figure}[htbp]
		 \begin{center}
		  \includegraphics[width=140mm]{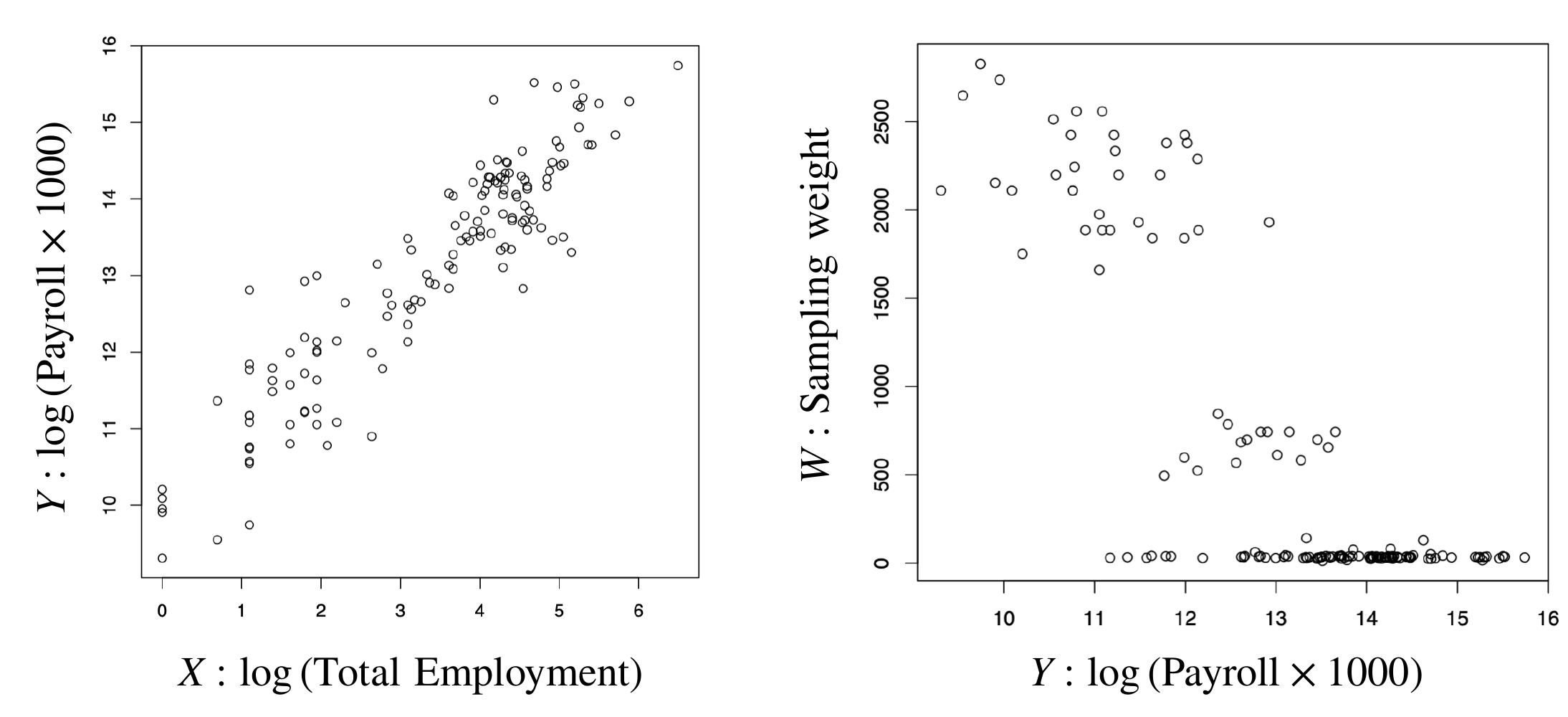}
		 \end{center}
		 \caption{Scatter plots of $(x_i, y_i)$ and $(y_i, w_i)$\; ($i = 1, \ldots, 142$).}
    \label{fig:3}
\end{figure}

This study analyzed a dataset mimicked for the 1999 Canadian Workplace and Employee Survey created in \citet{fuller09}, where the original survey is described in \citet{patak98}. The mimicked dataset is available from Example 6.3.3. in \citet{fuller09}. Let each $x_i$ and $y_i$ $(i=1, \ldots, 142)$ be the employment logarithm of 1999 total employment and the logarithm of 1999 payroll multiplied by 1000. According to \citet{fuller09}, the sampling mechanism was stratified sampling with three strata that depends on a variable highly correlated with $y$. In addition, the sampling mechanism in each stratum is simple random sampling with certain variance owing to adjustments for nonresponse. The visual relationships between $x$ and $y$, and $y$ and $w$ are shown in Figure \ref{fig:3}.

This study centers on estimating a parameter of the model $Y\mid x \sim N(a+b x, \sigma^2)$, where $\theta=(a, b, \sigma^2)^\top$. A working model \eqref{no_depend} was employed for weight distribution, as it is essentially simple random sampling in each stratum. Let the vectors of each mean and precision parameter in each stratum be $m=(m_1, m_2, m_3)^\top$ and $\phi = (\phi_1, \phi_2, \phi_3)^\top$ respectively, and let a model on the mixture proportion be
\begin{align}
&P(G=g\mid x,y; \alpha)
=\frac{\exp(\alpha_{g0}+\alpha_{g1}y)}{\sum_{h=1}^3 \exp(\alpha_{h0}+\alpha_{h1}y)}, \quad  (g=1,2,3) \label{mix_prop}
\end{align}
where $\alpha_{10}=\alpha_{11}=0$, and $\alpha=(\alpha_{20},\alpha_{21},\alpha_{30},\alpha_{31})^\top$.
The estimated values of $\alpha$, $m$, and $\phi$ were $\hat{\alpha}=(-0.884, -0.123, 25.899, -1.762)^\top$, $\hat{m}=(135.177, 1855.42, 34.421)^\top$, and $\hat{\phi}=$(9978.55, 3993.41, 471.57)$^\top$, respectively. Estimates of $\hat{\theta}=(\hat{a}, \hat{b},\hat{\sigma}^2)^\top$ using the six methods as in the numerical study are listed in Table \ref{tab:3}. Because the size of the finite population $N$ was unknown, and information of $x$ for units of $\delta = 0$ was unavailable, $\mathrm{Eff}^{00}_{\mathrm{reg}}$, $\mathrm{Eff}^{00}_{\mathrm{out}}$, and $\mathrm{Eff}^{00}_{\mathrm{out,DML}}$ were used for the proposed estimators. To stabilize the DML estimate, we set $L=1,000$ and calculated the median from 1,000 estimates. The standard error of the DML estimate is computed from the 200 bootstrap samples. As expected, estimates of HT and the two adaptive estimators $\mathrm{Eff}^{00}_{\mathrm{reg}}$ and $\mathrm{Eff}^{00}_{\mathrm{out}}$ were similar, while the CC and CML estimators differed. The primary reason for this difference between the CML and other consistent estimators is the misspecification of the sampling mechanism. However, regardless of any sampling mechanism misspecification, the proposed estimators with the parametric working model attained values similar to the HT estimator, and the estimated standard errors were smaller than those of the HT estimator. The DML estimate diverges slightly from the estimates with the parametric working model: the slope takes on larger values, while the intercept and conditional variance take smaller values. This difference is due to the fact that, unlike the parametric working model in \eqref{mix_prop}, we incorporate the information of $x$ into the sampling mechanism in the nonparametric working model, which could result in differing point estimates and reduced standard error.

\begin{table}[hbtp]
  \caption{Estimates for $\theta=(a, b, \sigma^2)$ with methods CC, HT, CML, $\mathrm{Eff}^{00}_\mathrm{reg}$, $\mathrm{Eff}^{00}_\mathrm{out}$, and $\mathrm{Eff}^{00}_\mathrm{out,DML}$. The values in parenthesis indicate the estimated standard error.}
  \label{tab:3}
  \centering
  \begin{tabular}{cccccccc}
    \hline \hline
                                           &  CC  & HT   & CML    & $\mathrm{Eff}^{00}_{\mathrm{reg}}$ & $\mathrm{Eff}^{00}_{\mathrm{out}}$ & $\mathrm{Eff}^{00}_{\mathrm{out, DML}}$ \\ \hline
     \multirow{2}{*}{$\hat{a}$}            &  13.082  & 12.889   & 12.827   & 12.895   & 12.886  & 13.011\\ 
                                           &  (0.049) & (0.113)  & (0.059)  &  (0.099) &  (0.094)& (0.064) \\ 
     \multirow{2}{*}{$\hat{b}$}            &  0.907   & 0.931    &  0.847    & 0.935   & 0.930   & 0.978 \\
                                           &  (0.033) & (0.054)   & (0.035)  & (0.047) & (0.044) & (0.041) \\ 
     \multirow{2}{*}{$\hat{\sigma}^2$}     &  0.316   & 0.299     & 0.309    & NA      & 0.298   & 0.254 \\
                                           &  (0.044) & (0.070)   & (0.042)  & NA      & (0.068) & (0.062)  \\ 
    \hline \hline
  \end{tabular}
\end{table}


\section{Conclusion}
\label{sec:7}

We have  proposed estimators for three different target parameters (i)--(iii) defined in \S \ref{sec:2.1} that can achieve the semiparametric efficiency bound under different settings: whether the size of the finite population $N$ is known; whether the information of $x$ for units of $\delta=0$ is available. The proposed adaptive estimators require specifying a model for the sampling weights.  We have considered two working models: a parametric working model based on beta regression, which is reasonable and computationally feasible; nonparametric working model by using double/debiased machine learning. The subsequent numerical study and the real data analysis show  that appropriate modeling of the sampling mechanism increased the efficiency of the estimators. 

However, we have implicitly assumed that there are no missing data while some data can be subject to missingness in practice. Developing semiparametric efficient estimator with missing data under informative sampling can be an important future research topic.  

\section*{Acknowledgements}
The authors are grateful for the very constructive comments of the two anonymous referees.

\section*{Funding}
The research of the first and the second authors were supported by a grant from MEXT Project for Seismology toward Research Innovation with Data of Earthquake (JPJ010217), the first author was supported by JSPS KAKENHI Grant (JP23H00466), the second author was supported by JSPS KAKENHI Grant (JP20K19756, JP20H00601, and JP23H03355), and the third author  was partially supported by a grant from US National Science Foundation (2242820)  and a grant from  the U.S. Department of Agriculture’s National Resources Inventory, Cooperative Agreement NR203A750023C006, Great Rivers CESU 68-3A75-18-504.

\appendix

\section{Technical Proofs}
\label{Appendix_B}

\noindent
\textbf{Proof of Theorem \ref{thm.1}(a)}. The orthogonal nuisance tangent space in the super population model is $\Lambda^{sp,\perp}=\{B U(X, Y; \theta): \mathrm{for~any~}B^{q\times q}~\mathrm{matrix
}\}$. Because $D^*_{\eff}\in\Lambda^{sp,\perp}$, there exists a unique $B^*_{\eff}$ such that $D^*_{\eff}=B^*_\eff U(\theta)$, where $B^*_\eff$ is a $q\times q$ invertible matrix. It follows from \eqref{eff_score} that the efficient score is 
$$S_\eff(X, Y,  W)=B^*_{\eff}\lllp \delta W U(\theta) + (1-\delta W)\frac{E\{(W-1)U(\theta)\mid X\}}{E(W-1\mid X)}\rrrp.$$
Because $B^*_\eff$ is invertible and the constant matrix does not affect efficiency, the efficient score is
$$S_\eff(X, Y,  W)= \delta W U(\theta) + (1-\delta W)\frac{E\{(W-1)U(\theta)\mid X\}}{E(W-1\mid X)}.$$

\noindent
\textbf{Proof of Theorem \ref{thm.1}(b)}. The nuisance tangent space in the super population model is expressed as $$\Lambda^{sp,\perp}=\{A(x)\varepsilon~\mathrm{for~any~}q\times 1\mathrm{~vector~}A(x)\},$$
where $\varepsilon=y-\mu(x;\theta)$. The projection of any $h\in\mathcal{H}$ onto the nuisance tangent space is
$\prod (h\mid \Lambda^{sp,\perp})=E(h\varepsilon\mid x)\varepsilon/V(x)$,
and the efficient score is $S^{sp}_\eff=\dot{\mu}(x; \theta)\varepsilon/V(x)$,
where $V(x)=E(\varepsilon^2\mid x)$. Further details can be found in Theorem 4.8 in Tsiatis (2006). Because $D^*_\eff \in \Lambda^{sp,\perp}$, there exists $A^*_\eff(x)$ such that $D^*_\eff =A^*_\eff(x)\varepsilon$. As $D^*_\eff$ satisfies \eqref{eff_eq}, for any vector $A(x)$,
\begin{align*}
    E\lllp A(X)\varepsilon \llp WA^*_\eff(X)\varepsilon - (W-1)C^*_\eff(X) - \frac{\dot{\mu}(X; \theta)}{V(x)}\varepsilon \rrp\rrrp=0,
\end{align*}
where $C^*_\eff(x)=E\{(W-1)A^*_{\eff}\varepsilon\mid x\}/E(W-1\mid x)$. Solving the above equation by equating to 0, the conditional expectation of the expression within square brackets yields
\begin{align*}
A^*_\eff(x)=\frac{1}{E(W\varepsilon^2\mid x)}\lllp E(W\varepsilon\mid x)C^*_\eff(x)+\dot{\mu}(x; \theta)\rrrp.
\end{align*}
The function $C^*_{\eff}(x)$ can be obtained by the relationship
\begin{align*}
    C^*_\eff(x) E(W-1\mid x) &= E\{(W-1)A_\eff^*(x)\varepsilon\mid x\}=C^*_\eff(x) \frac{\{E(W\varepsilon \mid x)\}^2}{E(W\varepsilon^2\mid x)}+
     \frac{E(W\varepsilon \mid x)}{E(W\varepsilon^2\mid x)} \dot{\mu}(x;\theta)
\end{align*}
Based on the Cauchy-Schwartz's inequality the following is obtained
\begin{align*}
\frac{\{E(W\varepsilon \mid x)\}^2}{E(W\varepsilon^2\mid x)}
&\leq \frac{E\{(W-1)\varepsilon^2\mid x\}}{E(W\varepsilon^2\mid x)} E(W-1\mid x)\\
&< E(W-1\mid x),
\end{align*}
that is, for all $x$,
$$E(W-1\mid x)- \frac{\{E(W\varepsilon \mid x)\}^2}{E(W\varepsilon^2\mid x)} \neq 0.$$
Thus, the explicit form of $C^*_\eff(x)$ is
\begin{align*}
C^*_\eff(x) 
&=\lllp E(W-1\mid x) - \frac{\{E(W\varepsilon \mid x)\}^2}{E(W\varepsilon^2\mid x)} \rrrp^{-1} \frac{E(W\varepsilon \mid x)}{E(W\varepsilon^2\mid x)} \dot{\mu}(x;\theta).
\end{align*}

\noindent
\textbf{Proof of Theorem \ref{thm.1}(c)}. The orthogonal nuisance tangent space in the super population model is $\Lambda^{sp,\perp}=\{h(x,y)\in\mathcal{H}\mid E(h(x,Y)\mid x)=0\}$ and the projection of any $h\in\mathcal{H}$ onto the nuisance tangent space is 
$\prod(h\mid \Lambda^{sp,\perp})=E(h\mid x,y)-E(h\mid x)$,
and the efficient score is $S^{sp}_\eff =S_\theta$. Let $D^*_\eff=D^*_\eff(x,y)$ such that $E(D^*_\eff\mid x)=0$. As $D^*_\eff$ satisfies \eqref{eff_eq}, for any $d(x,y)$ such that $E(d\mid x)=0$,
\begin{align*}
&E\lllp d(X,Y)\llp WD^*_\eff-(W-1)C^*_\eff(X) - S_\theta\rrp\rrrp\\
&=E\lllp d(X,Y)\llp E(W\mid X, Y)D^*_\eff-(E(W\mid X, Y)-1)C^*_\eff(X) - S_\theta\rrp\rrrp=0
\end{align*}
Because the above equation holds for any $d(x,y)\in\Lambda^{sp,\perp}$, the function inside the braces must be a function of $x$, for example, $k^*(x)$,
$$E(W\mid X, Y)D^*_\eff-(E(W\mid X, Y)-1)C^*_\eff(X) - S_\theta=k^*(X).$$
Using the function $k^*(x)$, $D^*_\eff$ is expressed as
\begin{align}
D^*_\eff=\bar{\pi}\llp S_\theta -\lp1-\frac{1}{\bar{\pi}}\rp C^*_\eff + k^*\rrp, \label{dstareff}
\end{align}
where $\bar{\pi}=\bar{\pi}(x,y)=1/E(W\mid x,y)$. 
By taking the conditional expectation with respect to $Y$ given $X=x$ in the above equation and utilizing the property $E(D^*_\eff\mid x) = 0$, we have
$$k^*(x)=- \frac{E(\bar{\pi} S_\theta\mid x)}{E(\bar{\pi}\mid x)}+\llp 1-\frac{1}{E(\bar{\pi}\mid x)}\rrp C^*_\eff(x).$$
Substituting the above $k^*(x)$ into \eqref{dstareff}, we have
\begin{align*}
      D^*_\eff = \bar{\pi}\llp S_\theta-\frac{E(\bar{\pi} S_\theta\mid x)}{E(\bar{\pi}\mid x)} +\lp \frac{1}{\bar{\pi}} - \frac{1}{E(\bar{\pi}\mid x)}\rp C^*_\eff(x) \rrp.
\end{align*}
Finally, we derive the explicit form of $C^*_\eff$.
    By simple algebra, we have
    \begin{align*}
    E\{(W-1)D^*_\eff\mid x\} &= E(WD^*_\eff\mid x)\\
    &= -\frac{E(\bar{\pi}S_\theta\mid x)}{E(\bar{\pi}\mid x)} +\llp E(W\mid x) - \frac{1}{E(\bar{\pi}\mid x)} \rrp C^*_\eff(x)
    \end{align*}
Hence, the relationship $C^*_{\eff}E(W-1\mid x) = E\{(W-1)D^*_{\eff}\mid x\}$ yields
\begin{align*}
C^*_{\eff}(x)\{E(W\mid x) - 1\} = -\frac{E(\bar{\pi}S_\theta\mid x)}{E(\bar{\pi}\mid x)} +\llp E(W\mid x) - \frac{1}{E(\bar{\pi}\mid x)} \rrp C^*_\eff(x),
\end{align*} 
which reduces to
\begin{align*}
C^*_{\eff}(x) = \frac{E(\bar{\pi} S_\theta\mid x)}{E(\bar{\pi}\mid x)-1}.
\end{align*}

\section{Regularity conditions}
\label{Appendix_A}

\noindent
In  \S4, parametric working models on $f(y\mid x; \xi)$ and $f(w\mid x, y; \psi)$ were assumed, where $\psi = (\phi^\top, \alpha^\top, \beta^\top,\gamma^\top)^\top$. Let compact parameter spaces of $\theta$, $\xi$, and $\psi$ be $\Theta$, $\xi$, and $\Psi$, respectively, the distribution function of $f(w\mid x,y;\psi)$ be $F_{x,y}(w;\psi)$, and $C=\{\psi'\in\Psi\mid F_{x,y}(w;\psi')=F_{x,y}(w;\psi)~\mathrm{a.s.}~\mathrm{for~all~continuity~points~}w\}$. Thus, any density function $f(w\mid x,y;\psi')$ for $\psi'\in C$ has the same distribution function as $f(w\mid x,y;\psi)$. Here, a quotient topological space $\tilde{\Psi}$ obtained from $\Psi$ by identifying parameters in $C$ to the same point was considered. This modification for the parameter space of $\Psi$ is required to avoid identification problems for mixture distributions \citep{redner81}.

Recall that the maximum likelihood estimators for $\xi\in\xi$ and $\psi\in\tilde{\Psi}$ are defined by
\begin{align*}
    \hat{\xi}_{\HT} = \argmax_{\xi\in\xi} \sum_{i=1}^N \delta_i W_i \log f(Y_i\mid X_i; \xi),\quad \hat{\psi} = \argmax_{\psi\in\tilde{\Psi}} \sum_{i=1}^N \delta_i f_s(O_i\mid X_i, Y_i; \psi), 
\end{align*}
where $f_s(o\mid x,y)$ is a mixture distribution of \eqref{odds_obs} with \eqref{mix_model}. Let $\theta$, $\xi$, $\psi$ be the true values of the parameters. Further, following the conditions B1--B6 are assumed, which are ordinary conditions for estimators of an estimating equation to have consistency and asymptotic normality \citep{newey94}.

\begin{itemize}
    \item[B1.] \label{A1} 
The finite population $(X_i, Y_i, W_i)\,(i=1, \dots, N)$ consists of independent and identically distributed realizations from a superpopulation. The sampled data are obtained through Poisson sampling from this finite population, where $\delta_i\mid (W_i=w_i) \sim Bernoulli(w_i^{-1})$.

    \item[B2.] \label{A2} 
The functions $m(x,y;\beta)$ and $\phi(x,y;\gamma)$ defined in \S 4.1 are identifiable, i.e., $m(x,y;\beta)=m(x,y;\beta')$ and $\phi(x,y;\gamma)=\phi(x,y;\gamma')$, with probability one under the sampled distribution, implies $\beta=\beta'$ and $\gamma=\gamma'$.

    \item[B3.] \label{A3}
    The maximum likelihood estimators $(\hat{\xi}_{\HT}^\top,~\hat{\psi}^\top)^\top$ converge to a vector of finite values $((\xi^*_{\HT})^\top, (\psi^*)^\top)^\top$ in probability and have asymptotic normality, where the asymptotic variance is nonsingular. If all the working models are correct, $((\xi^*_{\HT})^\top, (\psi^*)^\top)^\top = (\xi^\top, \psi^\top)^\top$.
    \item[B4.] \label{A4} 
    $\pd S_{\eff}(\theta,\xi_\HT, \psi)/\pd(\theta^\top,\xi_{\HT}^\top, \psi^\top)$ is continuous at $(\theta, \xi_{\HT}^*, \psi^*)$ with probability one, and there is a neighborhood $\Theta_{\mathcal{N}}\times \xi_{\HT, \mathcal{N}} \times \tilde{\Psi}_{\mathcal{N}}$ of $(\theta, \xi_{\HT}^*, \psi^*)$ such that 
$$\bigg\|E\bigg\{\sup_{(\theta, \xi_{\HT}, \psi)\in \Theta_{\mathcal{N}}\times \xi_{\HT, \mathcal{N}} \times \tilde{\Psi}_{\mathcal{N}}} \frac{\pd S_{\eff}(\theta,\xi_\HT, \psi)}{\pd(\theta^\top,\xi_{\HT}^\top, \psi^\top)}\bigg\}\bigg\|<\infty,$$
where for any matrix $A=(a_{ij})$, $\|A\|=\max_{i,j}|a_{ij}|$.
    \item[B5.]\label{A5} 
    $S_\eff(\theta,\xi_\HT, \psi)$ is continuously differentiable at each $(\theta, \xi_{\HT}^*, \psi^*)\in\Theta\times\xi\times\tilde{\Psi}$ with probability one, and there exists $d(X, Y, W)$ such that $\|S_\eff(\theta,\xi, \psi)\|\leq d(X, Y, W)$ for all $(\theta,\xi, \psi)\in\Theta\times\xi\times\tilde{\Psi}$ and $E\{d(X, Y, W)\}<\infty$.
    \item[B6.]\label{A6}
    $E\{ \pd S_\eff(\theta,\xi^*_{\HT}, \psi^*)/\pd(\theta^\top,\xi_{\HT}^\top, \psi^\top)\}$ is nonsingular at $(\phi, \xi^*_{\HT}, \psi^*)$.
\end{itemize}


\section{Efficent Scores in Setting 2, unknown $N$, and non-informative sampling}
\label{Appendix_D}
We provide the efficient scores in Setting 2 in Figure 1, when $N$ is unknown, and when the sampling mechanism is non-informative. We first consider the Setting 2, but we omit the proof because it is almost the same that for Theorem 1.

\begin{cor}
\label{cor.1}
Consider semiparametric estimators for each parameter (a)--(c) expressed in \S \ref{sec:2} under the Setting 2. Then, the efficient scores for the three estimators are
\begin{align}
\tilde{S}_\eff = \delta W \tilde{D}^*_\eff(\theta; X, Y) + (1-\delta W)\tilde{C}^*_\eff(\theta), \label{eff_est_eq2}
\end{align}
where $\tilde{D}^*_\eff$ and $\tilde{C}^*_\eff$ are different for the three parameters as follows:\\
\noindent
For $\theta$ defined through $E\{U(\theta; X, Y)\}=0$.
    \begin{align}
    \tilde{D}^*_\eff= U(\theta),\quad \tilde{C}^*_\eff = \frac{E\{(W-1)U(\theta)\}}{E(W-1)}.    \label{mean_eff2}
    \end{align}
\noindent
For $\theta$ in $\mu(x;\theta)=E(Y\mid x)$.
    \begin{align}
    \begin{split}
        \tilde{D}^*_\eff &= A^*_\eff (x)\{Y-\mu(x;\theta)\},\\
        \tilde{C}^*_\eff &= \lp E(W-1) - E\lllp \frac{\{E(W\varepsilon \mid X)\}^2}{E(W\varepsilon^2\mid X)} \rrrp \rp^{-1} E\lllp \frac{E(W\varepsilon \mid X)}{E(W\varepsilon^2\mid X)} \dot{\mu}(X;\theta) \rrrp, \label{reg_eff2}
    \end{split}
    \end{align}
        where
        $$A^*_\eff(x)=\frac{1}{E(W\varepsilon^2\mid x)}\lllp E(W\varepsilon\mid x)\tilde{C}^*_\eff+\dot{\mu}(x; \theta)\rrrp.$$
\noindent 
For $\theta$ in $f(y\mid x; \theta)$.
    \begin{align*}
      \tilde{D}^*_\eff &= \bar{\pi}\llp S_\theta-\frac{E(\bar{\pi} S_\theta\mid X;\theta)}{E(\bar{\pi}\mid X;\theta)}\rrp +\lp 1 - \frac{\bar{\pi}}{E(\bar{\pi}\mid X;\theta)}\rp \tilde{C}^*_\eff,\\
      \tilde{C}^*_{\eff} &= \frac{E\llp E(\bar{\pi} S_\theta\mid X;\theta)/E(\bar{\pi}\mid X;\theta)\rrp}{1-E\llp 1/E(\bar{\pi}\mid X;\theta)\rrp}.
    \end{align*}
The lower bound of the asymptotic variance is $\{E(\tilde{S}_\eff^{\otimes 2} )\}^{-1}$.
\end{cor}

Furthermore, the efficient score when $N$ is unknown can be derived. In this case, the augmented term is unnecessary in both Settings 1 and 2. 

\begin{cor}
\label{cor.2}
When $N$ is unknown, the efficient score is obtained by fixing $C^*_\eff(x)=0$ and $\tilde{C}^*_\eff=0$ in \eqref{eff_est_eq1} and \eqref{eff_est_eq2}; that is, the efficient scores are
\begin{align}
    S_{\eff} = \delta W D^*_{\eff} \label{c_zero},
\end{align}
where $D^*_{\eff}$ is different for the three parameters as follows:
    for $\theta$ defined through $E\{U(\theta; X, Y)\}=0$,  
    $D^*_\eff= U(\theta)$; for $\theta$ in $\mu(x;\theta)=E(Y\mid x)$, $D^*_\eff =\dot{\mu}(x; \theta)\{Y-\mu(x;\theta)\}/E(W\varepsilon^2\mid x)$;
    for $\theta$ in $f(y\mid x; \theta)$, $D^*_\eff = \bar{\pi}\{S_\theta - E(\bar{\pi}S_\theta\mid X;\theta)/E(\bar{\pi}\mid x;\theta)\}$.

\end{cor}

\noindent
\textbf{Proof of Corollary \ref{cor.2}}. By Proposition (A1.3) in \citet{rotnitzky97}. $\Lambda^\perp = \delta W D^*_\eff+A^{(2)}$, where $D^*_\eff\in\Lambda^{sp,\perp}$ and $A^{(2)}\in\Lambda^{(2)}$ and $\Lambda^{(2)}=\{h\in\mathcal{H}\mid E(h\mid X, Y, Z, W)=0\}$. When $N$ is unknown, there is no information on missing data, that is, for data $\delta_i=0$; thus, $\Lambda^{(2)}\equiv \{0\}$, which implies $C^*_\eff=0$. Then, the remaining proof is identical to that of Theorem \ref{thm.1}, and the expression of $D^*_\eff$ is the same except for the value of $C^*_\eff$ being fixed to 0.
\vspace{2ex}

Finally, the efficient scores when the sampling mechanism is non-informative, or $W$ is independent of $Y$ given $X$ is obtained. In this case, the augmented term is also unnecessary in both Settings 1 and 2, which can be shown through straightforward calculations. 

\begin{cor}
\label{cor.3}
When the sampling mechanism is non-informative in both Settings 1 and 2, the efficient score is given by \eqref{c_zero}, and the form of  $D^*_{\eff}$ is the same as that in Corollary S2 for parameters (a) and (b); however, for (c), $D^*_{\eff} = \bar{\pi}S_\theta$.
\end{cor}

\section{EM Algorithm for Mixture Working Models}
\label{Appendix_C}
We provide the E-step in the EM algorithm to estimate parameters given in \eqref{mix_model}. The M-step can be conducted by using the standard Newton method. With the models \eqref{mix_model}, the probability of sampled data belonging to a latent class is 
\begin{align*}
    P(G=g\mid x, y ,\delta =1) &\propto P(G=g\mid x, y)P(\delta=1\mid x, y, G=g)=p_gm_g.
\end{align*}
Then, the complete sampled log-likelihood is given by
\begin{align*}
&\ell_{\mathrm{com}}(\alpha, \beta, \phi)\\
&= \sum_{g=1}^H I(G=g)\llp \log P(G=g\mid x, y, \delta = 1)+\log f(w\mid x, y, G=g, \delta=1) \rrp\\
&= \sum_{g=1}^H I(G=g)\Big[ \log\llp p_g(\alpha_g)m_g(\beta_g, \phi_g)\rrp - \log \llp \sum_{h=1}^H p_h(\alpha_h)m_h(\beta_h, \phi_h)\rrp\\
&\quad +\log f(w\mid x, y, G=g, \delta=1; \beta, \phi) \Big].
\end{align*}
Note that $W-1\mid (x,y,G=g, \delta=1) \sim \mathrm{Beta'}(\{1-m_g(\beta_g)\}\phi_g, m_g(\beta_g)\phi_g+1)$. Then, the conditional expectation given observed data and estimated $(t-1)$th-step parameters is
\begin{align*}
    &E\{\ell_{\mathrm{com}}(\alpha, \beta, \phi)\mid w, x,y,\delta=1, \alpha^{(t-1)}, \beta^{(t-1)}, \phi^{(t-1)}\}\\
    &= \sum_{g=1}^H P(G=g\mid w,x,y,\delta=1;\alpha^{(t-1)},\beta^{(t-1)},\phi^{(t-1)})\\
    &\quad\times \left[ \log\llp p_g(\alpha_g)m_g(\beta_g, \phi_g)\rrp - \log \llp \sum_{h=1}^H p_h(\alpha_h)m_h(\beta_h, \phi_h)\rrp+\log f(w\mid x, y, G=g, \delta=1) \right],
\end{align*}
where
\begin{align*}
&P(G=g\mid w, x,y,\delta=1;\alpha^{(t-1)},\beta^{(t-1)},\phi^{(t-1)})\\
&=\frac{f(w\mid x, y, G=g, \delta = 1; \beta^{(t-1)}_g, \phi^{(t-1)}_g)p_g(\alpha^{(t-1)}_g)m_g(\beta^{(t-1)}_g, \phi^{(t-1)}_g)}{\sum_{h=1}^H f(w\mid x, y, G=h, \delta = 1; \beta^{(t-1)}_h, \phi^{(t-1)}_h)p_h(\alpha^{(t-1)}_h)m_h(\beta^{(t-1)}_h, \phi^{(t-1)}_h)}.    
\end{align*}

\bibliographystyle{imsart-nameyear}
\bibliography{ref}

\end{document}